\documentstyle[aps,draft,preprint]{revtex}

\def\be{\begin{equation}}
\def\bea{\begin{eqnarray}}
\def\bma{\begin{mathletters}}
\def\ee{\end{equation}}
\def\eea{\end{eqnarray}}
\def\ema{\end{mathletters}}

\begin{document}
\author{V. Vedral}
\title{Classical Correlations and Entanglement in Quantum Measurements
\footnote{This work is dedicated to Michael Vincent Vedral.}}
\address{Optics Section, Blackett Laboratory, Imperial College,
Prince Consort Road, London SW7 2BZ}
\date{\today}
\maketitle

\begin{abstract}
We analyze a quantum measurement where the apparatus is initially in a mixed state. We
show that the amount of information gained in a measurement is not equal to the amount
of entanglement between the system and the apparatus, but is instead equal to the
degree of classical correlations between the two. As a consequence, we derive an
uncertainty-like expression relating the information gain in the measurement and the
initial mixedness of the apparatus. Final entanglement between the environment and the
apparatus is also shown to be relevant for the efficiency of the measurement.
\end{abstract}

\vspace*{0.5cm}


Any measurement can be modeled as an establishment of correlations
between two random variables: one random variable represents the
values of the quantity pertaining to the system to be measured,
while the other random variable represents the states of the
apparatus used to measure the system \cite{Everett}. It is by
looking at the states of the apparatus, and discriminating them,
that we infer the states of the system. Looking at the apparatus,
of course, is another measurement process itself, which correlates
our mental states (presumably another random variable) with those
of the apparatus, so that indirectly we become correlated with the
system as well. It is at this point that we can say that we have
gained a certain amount of information about the system. This
description of the measurement process is true both in classical
and quantum physics. (Note that this way there is no more mystery
in the ``quantum state collapse" than there is in the
corresponding classical measurement).  The difference between the
two lies in the way we represent states of systems and the way we
represent their mutual interaction and evolution. Classically,
physical states of an $n$-dimensional system are vectors in a real
$n$ dimensional vector space whose elements are various
occupational probabilities for the states. The evolution of a
classical system is in general some stochastic map acting on this
vector space. Quantum mechanically, on the other hand, states are
in general represented using density matrices, while the evolution
is a completely positive, trace preserving transformation acting
on these matrices. Using this representation, classical physics
becomes a limiting case of quantum mechanics when the density
matrices are strictly diagonal in one and the same fixed basis and
the completely positive map then becomes the stochastic map.
Because of this fact, it is enough to analyze properties of
quantum systems and quantum evolutions and all the results are
automatically applicable to classical physics when we restrict
ourselves to the diagonal density operators only. A comprehensive
survey of major papers on quantum measurement can be found in
\cite{Wheeler} and the first fully quantum analysis was due to von
Neumann \cite{vonNeumann}.

In this letter we analyze a quantum measurement when the apparatus
is ``fuzzy", i.e. it is initially in a mixed state. Our approach
is entropic in character and is therefore closest in spirit to
that of Lindblad \cite{Lindblad}. We show that the amount of
information gained via the apparatus is proportional to the {\em
classical} correlations between the systems and the apparatus,
rather than the amount of entanglement between them. We then
derive an uncertainty-like expression which says that the sum of
the information gained in the measurement and the mixedness of the
apparatus (quantified by the von Neumann entropy
\cite{vonNeumann}) is bounded from the above by $\log N$, where
$N$ is the dimension of the apparatus. Our analysis builds on
recent results in quantum information theory concerning
quantification of entanglement in bi- \cite{VP98,VedralPRL} and
tripartite systems \cite{VedralJPhysA2} and separating classical
and quantum correlations \cite{VedralJPhysA}. Quantum information
theory has mainly been developed to understand computation and
communication supported by quantum systems, but this knowledge can
now be applied back to quantum mechanics to study its foundations
from a new perspective.

We first review the existing measures of entangled and total
correlations \cite{Vrev}. In classical information theory, the
Shannon entropy, $H(X)\equiv H(p)=-\sum _{i}p_{i}\log p_{i}$, is
used to quantify the information in a random variable, $X$, that
contains states $x_{i}$ with probabilities $p_{i}$ \cite{Shannon}.
In the quantum context, the results of a projective measurement
$\{E_{y}\}$ on a state represented by a density matrix, $\rho$,
comprise a probability distribution $p_{y}=Tr(E_{y}\rho)$. Von
Neumann showed that the lowest entropy of any of these probability
distributions generated from the state $\rho$ was achieved by the
probability distribution composed of the eigenvalues of the state,
$\lambda=\{\lambda_{i}\}$ \cite{vonNeumann}. This probability
distribution would arise from a projective measurement onto the
state's eigenvectors. The Von Neumann entropy is then given by
$S(\rho)=-Tr(\rho\log\rho)=H(\lambda)$. The quantum relative
entropy of a state $\rho$ with respect to another state $\sigma$
is defined as $S(\rho ||\sigma)=-S(\rho)-Tr(\rho\log\sigma)$. The
joint entropy $S(\rho_{AB})$ for a composite system $\rho_{AB}$
with two subsystems $A$ and $B$ is given by
$S(\rho_{AB})=-Tr(\rho_{AB}\log\rho_{AB})$ and the Von Neumann
mutual information between the two subsystems is defined as
$I(\rho_{AB})=S(\rho_{A})+S(\rho_{B})-S(\rho_{AB})$. The mutual
information is the relative entropy between $\rho_{AB}$ and
$\rho_{A}\otimes\rho_{B}$. The mutual information is used to
measure the total correlations between the two subsystems of a
bipartite quantum system. The entanglement of a bipartite quantum
state $\rho_{AB}$ may be measured by how distinguishable it is
from the `nearest' separable state, as measured by the relative
entropy. Relative entropy of entanglement, defined as
\[
E_{RE}(\rho_{AB})=\min_{\sigma_{AB}\in D}S(\rho_{AB}||\sigma_{AB})
\]
has been shown to be a useful measure of entanglement ($D$ is the
set of all separable or disentangled states) \cite{VP98}. Note
that $E_{RE}(\rho_{AB})\leq I(\rho_{AB})$, by definition of
$E_{RE}(\rho_{AB})$, since the mutual information is also the
relative entropy between $\rho_{AB}$ and a completely disentangled
state. There are many other ways of measuring the entanglement of
a bipartite quantum state \cite{Vrev}, but they can all be unified
under the formalism of relative entropy \cite{VedralPRL}. Another
advantage of relative entropy is that it can be generalized to any
number of subsystems, a property that will be very useful in
understanding the measurement process when the environment is also
present. I will drop the subscript ``RE" to denote the relative
entropy of entanglement as this is the only measure that will be
used throughout. As a general comment we stress that all the
measures used here are entropic in nature, which means that they
are generally attainable only asymptotically. The advantage of
using entropic measures is that our results will be universally
valid, although they will almost always be overestimates in the
finite-case scenario.

Recently we have suggested that correlations in a state $\rho_{AB}$ can also be split
into two parts, the quantum and the classical part \cite{VedralJPhysA} (see also
\cite{Zurek,Divincenzo} for alternative approaches). The classical part is seen as the
amount of information about one subsystem, say $A$, that can be obtained by performing
a measurement on the other subsystem, $B$. The resulting measure is the difference
between the initial and the residual entropy \cite{VedralJPhysA}:
\begin{equation}
C_{B}(\rho_{AB})=\max_{B_{i}^{\dagger}B_{i}}S(\rho_{A})-\sum_{i}p_{i}
S(\rho_{A}^{i})\label{eq:cb}
\end{equation}
where $B_{i}^{\dagger}B_{i}$ is a Positive Operator Valued Measure
performed on the subsystem $B$ and
$\rho_{A}^{i}=tr_{B}(B_{i}\rho_{AB}B_{i}^{\dagger})/tr_{AB}
(B_{i}\rho_{AB}B_{i}^{\dagger})$ is the remaining state of $A$
after obtaining the outcome $i$ on $B$. Alternatively,
$C_{A}(\rho_{AB})=\max_{A_{i}^{\dagger}A_{i}}S(\rho_{B})-\sum_{i}p_{i}
S(\rho_{B}^{i})$ if the measurement is performed on subsystem $A$
instead of on $B$. Clearly $C_{A}(\rho_{AB})=C_{B}(\rho_{AB})$ for
all states $\rho_{AB}$ such that $S(\rho_{A})=S(\rho_{B})$ (e.g.
pure states). It remains an open question whether this is true in
general (but this will not affect our measurement analysis as the
apparatus is always measured to infer the state of the system and
never the other way round). This measure is a natural
generalisation of the classical mutual information, which is the
difference in uncertainty about the subsystem $B$ ($A$) before and
after a measurement on the correlated subsystem $A$ ($B$). Note
the similarity of the definition to the Holevo bound which
measures the capacity of quantum states for classical
communication \cite{Holevo}. The following example provides an
illustration of this and will be the key to our discussion of the
quantum measurement. Consider a bipartite separable state of the
form $\rho_{AB}=\sum_{i}p_{i}\left| i\right\rangle \langle
i|_{A}\otimes\rho _{B}^{i}$, where $\{\left| i\right\rangle \}$
are orthonormal states of subsystem $A$. Clearly the entanglement
of this state is zero. The best measurement that Alice can make to
gain information about Bob's subsystem is a projective measurement
onto the states $\{\left| i\right\rangle \}$ of subsystem $A$.
Therefore the classical correlations are given by
\[
C_{A}(\rho_{AB})=S(\rho_{B})-\sum_{i}p_{i}S(\rho_{B}^{i})
\]
which is, for this state, equal to the mutual information $I(\rho_{AB})$. This is to
be expected since there are no entangled correlations and so the total correlations
between $A$ and $B$ should be equal to the classical correlations. This measure of
classical correlations has other important properties such as $C(\rho_{AB})=0$ if and
only if $\rho_{AB}=\rho_{A}\otimes\rho_{B}$; it is also invariant under local unitary
transformations and non-increasing under any general local operations
\cite{VedralJPhysA}.

Let us now introduce the general framework for a quantum
measurement (for a special case see \cite{Bose}). We have a system
in the state $|\Psi\rangle = \sum_i a_i |i\rangle$, and an
apparatus in the state $\rho = \sum_i r_i |r_i\rangle \langle
r_i|$ in the eigenbasis. The purpose of a measurement is to
correlate the system with the apparatus so that we can extract the
information about the state $|j\rangle$ of the system. In a
perfect measurement, by looking at the apparatus we can
unambiguously identify the state of the system. Therefore, when
the system is in the state $|j\rangle$ we would like the apparatus
to be in the state $\rho_j$, such that $\rho_i \rho_j = 0$, i.e.
different states of the apparatus lie in orthogonal subspaces and
can be discriminated with a unit efficiency. If this condition is
not fulfilled, which is frequently the case, then the measurement
is imperfect and the amount of information obtained is not maximal
(this is what defines an ``imperfect measurement"). We now compute
the amount of information gained in general and show that it is
more appropriately identified with the classical rather than
quantum correlations between the system and the apparatus. Suppose
that the measurement transformation is given by a unitary
operator, $U$, acting on both the system and the apparatus, such
that
\[
U (\rho \otimes |i\rangle\langle j|)U^{\dagger} = \rho_{ij}
\otimes |i\rangle\langle j|
\]
where we assume that the measurement transformation acts such that
the state $|r_k\rangle |l\rangle$ of the apparatus and the system
respectively is transformed into the state $|\tilde{r}_{kl}\rangle
|l\rangle$, such that the states of the apparatus corresponding to
different system states are orthogonal $\langle
\tilde{r}_{ij}|\tilde{r}_{ik}\rangle = \delta_{jk}$. This
particular interaction is chosen so that in the special case of
the pure apparatus we obtain von Neumman's (and Everett's)
analysis. We see that the measurement is such that the new
apparatus state depends on the state of the system. This is
exactly how correlations between the two are established. Then,
the initial state is transformed into
\[
\rho_f = \sum_{ij} a_i a^*_j \rho_{ij} \otimes |i\rangle\langle j|
= \sum_i |a_i|^2 \rho_{ii} \otimes |i\rangle\langle i| +
\sum_{i\neq j} a_i a^*_j \rho_{ij} \otimes |i\rangle\langle j|
\]
The first term on the right hand side indicates how much information this measurement
carries. We will now measure the apparatus and try to distinguish the states
$\rho_{ii}$ to the best of our ability. Once we confirm that the apparatus is in the
state $\rho_{jj}$, then we can infer that the system is in the state $|j\rangle$. The
amount of information about the state of the apparatus (and hence the state of the
system), $I_m$, is given by the well-known Holevo bound \cite{Holevo}:
\begin{equation}
I_m = S(\sum_i |a_i|^2 \rho_{ii}) - \sum_i |a_i|^2 S(\rho_{ii})
\end{equation}
As we have seen, this quantity is also equal to the amount of
classical correlations between the system and the apparatus in the
state $\rho^{\prime}_f = \sum_i |a_i|^2 \rho_{ii} \otimes
|i\rangle\langle i|$, which is, in this case, the same as the von
Neumann mutual information between the two. Note that this state
is only classically correlated and there is no entanglement
involved. The amount of entanglement in the state $\rho_f$, on the
other hand, will in general be non-zero. This may be difficult to
calculate. However, we can provide lower and upper bounds. The
lower bound on the entanglement between the system and the
apparatus is
\begin{equation}
E(\rho_f) \geq S(\sum_i |a_i|^2 \rho_{ii}) - S(\rho_f) = S(\sum_i
|a_i|^2 \rho_{ii}) - S(\rho) = I_m
\end{equation}
(Note that here $S(\rho) = S(\rho_{ii}$ for all $i$ by definition
of measurement interaction). Therefore, the entanglement between
the system and the apparatus is larger than or equal to the
classical correlations between the two which quantify the amount
of information that the measurement carries. So, this shows that
the information in a quantum measurement is correctly identified
with the classical correlations between the apparatus and the
system rather than the entanglement or the mutual information
between the two in the final state, $\rho_f$. Only in the limiting
case of the pure apparatus do we have that the amount of
information in the measurement is equal to the entanglement, which
becomes the same as the classical correlations, while the sum of
the quantum and classical correlations is then equal to the mutual
information in the state. We stress again that the amount of
information in a single measurement will in general be less than
this quantity, which can usually only be reached in the asymptotic
limit.

We can recast this relationship in the form of an ``uncertainty relation" between the
initial mixedness of the apparatus and the amount of information gained. So, from the
fact that $I_m = S(\sum_i |a_i|^2 \rho_{ii}) - S(\rho)$, we have that
\begin{equation}
I_m + S(\rho) = S(\sum_i |a_i|^2 \rho_{ii}) \leq \log N
\end{equation}
where $N$ is the dimension of the apparatus. Thus we see that the sum of the initial
mixedness of the apparatus and the amount of information the measurement obtains is
always smaller than a given fixed value: the larger $S(\rho)$, the smaller $I_m$. When
$\rho$ is maximally mixed (and therefore $S(\rho) = \log N$), then no information can
be extracted from the measurement. Note that this relation is different to the usual
``information versus disturbance" law in a quantum measurement as well as to the usual
entropic uncertainty relations of incompatible observables. Every measurement that
extracts information from a quantum system also disturbs the state, and without this
disturbance there would be no information gain possible. The initial state of the
system in our above scenario was $\sum_i a_i |i\rangle$, while the final state is a
mixture of the form $\sum_i |a_i|^2 |i\rangle \langle i|$. The disturbance to the
state can be measured as a distance between the final and the initial state. We choose
the relative entropy to quantify this difference. So, while the information in the
measurement is given by $I_m$, the disturbance is
\[
D = S(|\Psi\rangle \langle \Psi |||\sum_i |a_i|^2 |i\rangle \langle i|) = -\sum_i
|a_i|^2 \log |a_i|^2
\]
which is the same as the maximum amount of information possible
from this measurement. So, the measurement described above always
maximally disturbs the state, and the reason why this does not
lead to the maximum information gain is because the apparatus
state is mixed. The system could be disturbed less by adjusting
the overlap between the states of the apparatus
$|\tilde{r}_{ij}\rangle$, so that they are not orthogonal to each
other. In general we can require that $\langle
\tilde{r}_{ij}|\tilde{r}_{ik}\rangle = a_{jk}$, such that
$|a_{jk}|<1$. We will not treat this case here: it is
mathematically more demanding, but does not illuminate the
measurement issue any better. Note also that a question may be
raised as to why we consider the interaction between the apparatus
and system to be unitary and not of a more general kind (a
completely positive map as in, for example \cite{Vrev}). The
reason is that any such interaction can be represented by a
unitary transformation \cite{Vrev} and our analysis then also
applies (although the resulting effective measurement would in
general be less efficient than the one performed unitarily).

In order to show that some form of entanglement is still important
(albeit not the one between the system and the apparatus) we
revisit the same measurement scenario, but from the ``higher
Hilbert space perspective". This is done by adding the environment
to the apparatus so that the joint state is pure,
$|\Psi_{EA}\rangle$. We briefly note that our treatment differs
from the usual ``environment induced collapse" and decoherence as
in, for example, \cite{Zurekdeco,Partovi}. In our case, the
environment is not there to cause the disappearance of
entanglement between the system and the apparatus, but is there to
purify the initally mixed state of the apparatus. We do not have
any state reduction or collapse, but just different ways in which
we can express the classical correlations between the system and
the apparatus. The measurement transformation is now given by
\[
|\Psi_{EA}\rangle \otimes \sum_i a_i |i\rangle \longrightarrow
\sum_i a_i |\Psi^i_{EA}\rangle |i\rangle
\]
where $|\Psi_{EA}\rangle = \sum_i \sqrt{r_i} |e_i\rangle |r_i\rangle$ and
$|e_i\rangle$ is an orthonormal basis for the environmental states. We see that when
the environment is traced out, the state of the apparatus is equal to $\rho$. Now, the
measurement implements a unitary transformation so that each of the states of the
apparatus changes according to which state of the system it interacts with. Therefore
we see that the $i$-th state of the environment and the apparatus after the
interaction is given by $|\Psi^i_{EA}\rangle = \sum_j \sqrt{r_j} |e_j \rangle
|\tilde{r}_{ji}\rangle $. To make a link with the first picture of the measurement, we
trace out the environment to obtain: $ \rho_{A'S'} = \sum_{ij} a_i a^*_j (\sum_k
\langle e_k|\Psi^i_{EA}\rangle \langle \Psi^j_{EA}|e_k\rangle) \otimes
|i\rangle\langle j|$ and, thus, the quantity in brackets can be identified with
$\rho_{ij} = \sum_k r_k |\tilde{r}_{ki}\rangle \langle \tilde{r}_{kj}|$. Therefore,
since we have no access to the environment, our task is to discriminate the states
$\rho_{ii}$, and therefore identify the corresponding states $|i\rangle$ of the system
and this was done in the previous analysis. If, on the other hand, we had access to
the environment, the measurement could be perfect.

We first apply entropic considerations to the ``environment-apparatus-system"
tripartite state. The initial and the final entropy of the environment are the same as
its state remains unchanged, and this value is the same as the initial entropy of the
apparatus, $S(\rho)$. As we have seen, this is an important quantity as it determines
how much information can be extracted from a measurement: the more mixed the initial
state of the apparatus, the less information can be extracted. If the initial state is
maximally mixed (say it is a thermal state with an arbitrarily high temperature), then
there can be no information gain during the measurement. The initial entropy of the
apparatus is also equal to the entropy of the system and the apparatus after the
measurement, $S(\rho_{A'S'}) \equiv S(\rho_f)$, as well as the amount of entanglement
between the environment and the system and the apparatus together, $E_{E:(A'S')}$,
after the measurement. The entanglement and the mutual information between the
environment and the apparatus after the measurement are always less than or equal to
their value before the measurement (since the systems becomes correlated to the
apparatus during the measurement).

We now use the recently derived three-party entanglement bounds to
provide further constraints on the measurement. For any pure
tripartite states $\sigma_{ABC}$ we have that
\cite{VedralJPhysA2}:
\begin{eqnarray}
&  & \max\{ E(\sigma_{AB}) + S(\sigma_{C}),E(\sigma_{AC}) +
S(\sigma_{B}),E(\sigma_{BC}) + S(\sigma_{A})\} \le
E(\sigma_{ABC}) \nonumber \\
&  & \le   \min \{S(\sigma_A) + S(\sigma_B), S(\sigma_A)+ S(\sigma_C), S(\sigma_B) +
S(\sigma_C) \label{ineq}\}
\end{eqnarray}
Applying this to our measurement scenario we obtain that
\[
S(\rho) \le E_{E:A^{\prime}:S^{\prime}} -
E_{A^{\prime}:S^{\prime}}
\]
where the subscripts $E,A,S$ indicate the environment, the apparatus and the system
respectively. The primes on the subscripts indicate states after the measurement. We
see that the closer the tripartite entanglement to the entanglement between the system
and the apparatus (with the environment disentangled), the more efficient the
measurement.  We immediately conclude that the necessary condition for the equality
between the two entanglements is that the initial entropy of the apparatus is zero. It
should be remembered, however, that the measurement can still be perfect even though
the apparatus is not pure and this is because the relevant quantity is the classical
correlations between the system and the apparatus and not their entanglement. In that
context we can also derive from the inequality in (\ref{ineq}) that $E_{E:A^{\prime}}
+ S(\rho_{S^{\prime}}) \le E_{E:A^{\prime}:S^{\prime}}$, so that
\[
E_{E:A^{\prime}} + I_m \le E_{E:A^{\prime}:S^{\prime}} -
S(\rho)\le S(\rho_A^{\prime})
\]
Thus, the sum of the information from the measurement and the
final entanglement between the environment and the apparatus is
limited by the final entropy of the apparatus and therefore by
$\log N$. Again we see that the larger the information we want,
the smaller the entanglement with the environment and the
apparatus will be. So, in fact, for the measurement to be
efficient we wish the environment {\em not} to become entangled
with the apparatus to a large extent (after the measurement). We
should mention at the end that our example is somewhat simplified
in that the environment will not, in reality, be passive
throughout the process. It would instead interact with both the
system and the apparatus making the measurement even less
effective, although all the above results would still apply.

In this letter we have analyzed the information gained in a quantum measurement when
the apparatus used to extract this information is initially in a mixed state. This is
a realistic scenario as the apparatus is usually assumed to be macroscopic and it is
consequently in thermal equilibrium with its own environment. We have shown that the
amount of information is correctly identified with the amount of classical
correlations between the system and the apparatus after their correlation is
established and derived an entropic uncertainty relation between this amount and the
mixedness of the initial state. Further light on quantum measurement was then shed by
purifying the apparatus and including its own environment in the analysis. Among open
problems highlighted by this work are to extend the analysis to non-orthogonal states
of the apparatus in the measurement transformation and to prove that the information
gain is symmetric between the system and the apparatus.

\vspace*{0.1cm}

\noindent {\bf Acknowledgments.} This work is funded by
Engineering and Physical Sciences Research Council, the European
grant EQUIP and Hewlett-Packard. The great hospitality of the
staff at the Maternity Ward of the Raigmore Hospital in Inverness,
Scotland is acknowledged where part of this work was completed.


\bigskip

\bigskip


\bigskip

\begin{references}
%
\bibitem{Everett} H. Everett, Rev. Mod. Phys. 29, 454 (1957); see also H. Everett,
editors B. S. DeWitt, R. Neill Graham eds., {\em The many-worlds Interpretation of
Quantum Mechanics}, (Princeton Series in Physics, Princeton University Press (1973)).
%
\bibitem{Wheeler} J. A. Wheeler, W. H. Zurek eds., {\em Quantum Theory and
Measurement}, (Princeton Series in Physics, Princeton University
Press (1983)).
%
\bibitem{vonNeumann} J. von Neumann, {\em Mathematical Foundations of Quantum
Mechanics}, (Princeton University Press, New York, 1955).
%
\bibitem{Lindblad} G. Lindblad, Comm. Math. Phys. {\bf 33} 305 (1973).
%
\bibitem{VP98} V. Vedral and M. B. Plenio, Phys. Rev. A, {\bf 57}, 1619 (1998).
%
\bibitem{VedralPRL} L. Henderson and V. Vedral, Phys. Rev. Lett. {\bf 84}, 2263 (2000).
%
\bibitem{VedralJPhysA2} M. B. Plenio and V. Vedral, J. Phys. A {\bf 34}, 6997 (2001).
%
\bibitem{VedralJPhysA} L. Henderson and V. Vedral, J. Phys. A {\bf 34}, 6899 (2001).
%
\bibitem{Vrev} V. Vedral, Rev. Mod. Phys. {\bf 74}, 297 (2002).
%
\bibitem{Shannon} E. Shannon and W. Weaver, {\em The Mathematical Theory of Communication},
(University of Illinois Press, Urbana, IL (1949).
%
\bibitem{Zurek} H. Ollivier and W. H. Zurek, Phys. Rev. Lett. {\bf 88}, 017901 (2002).
%
\bibitem{Divincenzo} B. M. Terhal, M. Horodecki, D. W. Leung and D. P. DiVincenzo,
lanl e-print no. 0202044 (2002).
%
\bibitem{Bose} S. Bose and V. Vedral, lanl e-print no. 0004016 (2000).
%
\bibitem{Holevo} A.S. Holevo, Probl. Peredachi Info. {\bf 9}, 177 (1973).
%
\bibitem{Zurekdeco} W. H. Zurek, Phys. Rev. D {\bf 24}, 1516 (1981).
%
\bibitem{Partovi} M. H. Partovi, Phys. Lett. A {\bf 137}, 445 (1985).

\end{references}
\end{document}